# Heralded linear optical quantum Fredkin gate based on one auxiliary qubit and one single photon detector[*]

Zhu Chang-Hua[†], Cao Xin, Quan Dong-Xiao, Pei Chang-Xing

State Key Laboratory of Integrated Services Networks, Xidian University, Xi'an 710071, China

**Abstract:** Linear optical quantum Fredkin gate can be applied to quantum computing and quantum multi-user communication network. In the existing linear optical scheme, two single photon detectors (SPDs) are used to heralding the success of the quantum Fredkin gate while they have no photon count. But analysis results show that for non-perfect SPD the lower the detector efficiency, the higher the heralded success rate by this scheme. We propose an improved linear optical quantum Fredkin gate by designing a new heralding scheme with an auxiliary qubit and only one SPD, in which the higher the detection efficiency of the heralding detector, the higher the success rate of the gate. The new heralding scheme can also work efficiently under non-ideal single photon source. Based on this quantum Fredkin gate, large-scale quantum switching networks can be built. As an example, a quantum Benes network in which only one SPD is used is shown.

# 1 Introduction

Quantum information science aims to apply the principle of quantum mechanism and unique quantum mechanical effects to enhance the performance in the transmission, processing and storage of information. Single photon is one of the carriers of quantum information for its low decoherence rate and easier manipulation in single photon level. Single photon has been widely used in quantum communication (e.g. quantum key distribution[1]), quantum computing (e.g. quantum logical gate[2][3]), quantum

[*] This work was supported by the National Natural Science Foundation of China No.61372076 and No. 61301171, the 111 Project no.B08038, and the Fundamental Research Funds for the Central Universities no. K5051301059 and K5051201021.
[†] Corresponding author. chhzhu@xidian.edu.cn

metrology[4][5], lithography[6], etc. In multi-user quantum communication network and quantum computing, quantum Fredkin gate plays an important role. By using quantum Fredkin gate multiple users can be interconnected to build a large-scale quantum communication network. With the development of integrated optics, large-scale optical quantum circuit or optical quantum computer, in which quantum Fredkin gate is one of the important gates, can be constructed[7][8].

Quantum Fredkin gate is also known as controlled swap gate, that is, if the control bit is in state $|1\rangle$, then the two target qubits swap their states and otherwise, they remain in their initial states if the control bit is in the state $|0\rangle$. Quantum Fredkin gate can be built by three CNOT gates[9], or can be built directly[10]. There are two types of optical methods to implement Quantum Fredkin gate. One is linear optical method by which the Fredkin gate is created by linear optical components; the other is nonlinear optical method by which nonlinear effect of optical components is used.

The scheme and experiments with linear optical method are developed quickly. In [11], Fiurášek proposed a quantum Fredkin gate by simulating the Kerr medium in Milburn's optical Fredkin gate[12] with linear interferometer and controlled phase gates. It needs six ancillary photons with the success probability $4.1 \times 10^{-3}$. Gong, Guo and Ralph constructed a heralded Fredkin gate by using four CNOT gates and linear optical components[10], they also simplified the method to a post-selection one which operates in the coincidence basis and required ancillary photons in maximally entangled state. In this method, several interferometers would have to be stabilized simultaneously, and the success of the Fredkin gate is conditioned under three-fold coincidence detection in the output modes. Fiurášek proposed a linear optical heralded quantum Fredkin gate based on partial SWAP gate and controlled-z gates[13], the suggested setup requires a Mach-Zehnder interferometer and six auxiliary photons forming maximally entangled EPR pairs. Then they simplified the design and gave a scheme operating in the coincidence basis which required one entangled EPR pair and four partially polarizing beam splitters. In the post-selected scheme (operating in coincidence basis)[10][13], the photons at the outputs are consumed and are not available for further processing unless quantum no-demolition detector are used.

The nonlinear optical quantum Fredkin gate is based on nonlinear effect. Huang and Kumar proposed an interaction-free quantum optical Fredkin gate using $\chi^{(2)}$ LiNbO$_3$ microdisk cavity based on the optically-induced quantum-zeno effect[14]. In this scheme photon loss is minimized due to elimination of the direct physical coupling between signal and the pump waves. Quantum noise level is also lowerd owing the weak background scattering in $\chi^{(2)}$ crystalline cavity. Recently, Hu, Huang and Kumar proposed a method to overcome the quantum fluctuations of pump waves by exploiting stimulated Raman in the implementation of quantum optical Fredkin gate[15]. In the nonlinear scheme, a pump wave is required. Lin proposed a scheme, with the success rate 1/133, which is based on the cross-Kerr effect between the photons and the ancillary state[16]. Shao, Chen and Zhang proposed a scheme which is based on the dispersive atom-cavity interaction. By modulating the cavity frequency and the atomic transition frequency, it can obtain an effective form of nonlinear interaction between photons in the three-mode cavity[17]. In addition, Wang, Zhang and Zhu also proposed a scheme to implement fermionic quantum Fredkin gate for spin qubits with the aid of charge detection[18].

In this paper, we focus on linear optical quantum Fredkin gate. In the existing heralding scheme[10][13], several flag detectors are used to indicate the success of the quantum Fredkin gate while they have no photon count. But our analysis results show that for non-perfect single photon detector (SPD) the lower the detector efficiency, the higher the ratio of successful heralding, see the section 2. Although the result whether the quantum fredkin gate is successful or not can be determined by coincidence count in the inputs and outputs, this heralding way may not be a good choice for on-line testing of each Fredkin gate of a large-scale quantum network. In this paper, we modify the heralding way in which the higher the detection efficiency of the heralding detector is, the higher the success rate of the quantum Fredkin gate is. Compared with the heralded scheme proposed in [10], our scheme requires only one detector and the success rate of which is increasing with increasing detector efficiency. Compared with the heralded scheme in [13], our scheme doesn't require interferometer and entangled auxiliary photons. We also show the large quantum switching network can be built with this Fredkin gate by using only one SPD, e.g. quantum Benes network.

The rest of the paper is organized as follows: in section 2, a heralded linear optical quantum Fredkin gate is presented and its success rate under low-efficient detector is analyzed. The success rate for input pulses with non-ideal single photon source (SPS) is also analyzed. In section 3, a quantum Fredkin gate with a new heralding scheme is proposed and its success rate is discussed. Based on the quantum Fredkin gate, we build a quantum Benes network by using only one SPD. In section 4, we conclude the paper.

## 2 Heralded linear optical quantum Fredkin gate

Here qubit is implemented by the photon with special polarization direction. The state $|0\rangle$ is prepared by horizontal polarization of a photon, denoted by $|H\rangle$ and the state $|1\rangle$ is prepared by vertical polarization direction of a photon, denoted by $|V\rangle$. So the operation for optical quantum bit is converted to the operation of the polarization of photon. The two input quantum bits of a Fredkin gate are $|\psi_{a_1}\rangle = \beta_{1h}|H\rangle_{a_1} + \beta_{1v}|V\rangle_{a_1}$ and $|\psi_{a_2}\rangle = \beta_{2h}|H\rangle_{a_2} + \beta_{2v}|V\rangle_{a_2}$, where $\beta_{1h}$, $\beta_{1v}$, $\beta_{2h}$ and $\beta_{2v}$ are arbitrary complexes, and $|\beta_{1h}|^2 + |\beta_{1v}|^2 = 1$, $|\beta_{2h}|^2 + |\beta_{2v}|^2 = 1$, $a_1$ and $a_2$ denote different space mode. So the joint state of the two input photons is

$$|\psi_{in}\rangle = |\psi_{a_1}\rangle|\psi_{a_2}\rangle = \left(\beta_{1h}|H\rangle_{a_1} + \beta_{1v}|V\rangle_{a_1}\right)\left(\beta_{2h}|H\rangle_{a_2} + \beta_{2v}|V\rangle_{a_2}\right) \quad (1)$$
$$= \beta_{1h}\beta_{2h}|H\rangle_{a_1}|H\rangle_{a_2} + \beta_{1h}\beta_{2v}|H\rangle_{a_1}|V\rangle_{a_2} + \beta_{1v}\beta_{2h}|V\rangle_{a_1}|H\rangle_{a_2} + \beta_{1v}\beta_{2v}|V\rangle_{a_1}|V\rangle_{a_2}$$

**2.1 The success rate of heralded linear optical quantum Fredkin gate under ideal SPS**

A typical quantum Fredkin gate is shown as Fig. 1 [10]. It can be in "Cross" state or "Through" state controlled by four controlled-Not (CNOT) gates. The "Cross" state means that the two input qubits are swapped, the qubit at $b_{1out}$ is in state $|\psi_{a_2}\rangle$ and the qubit at $b_{2out}$ is in state $|\psi_{a_1}\rangle$. The "Through" state means that after the two input qubits are passed through the qubit at $b_{1out}$ is in state $|\psi_{a_1}\rangle$ and the qubit at $b_{2out}$ is in state $|\psi_{a_2}\rangle$.

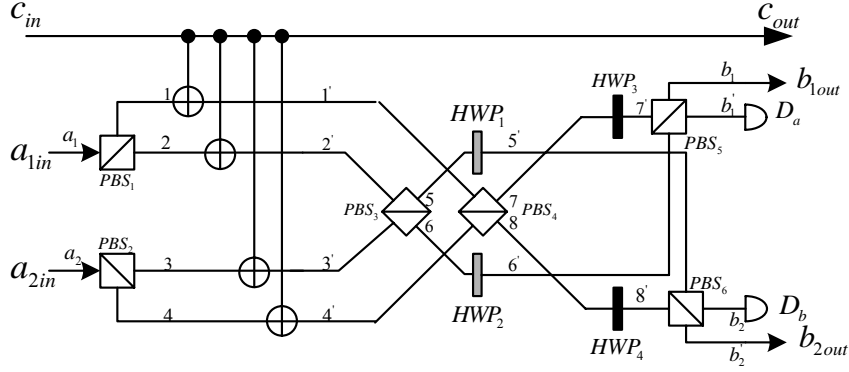

Fig.1 A linear optical implementation of quantum Fredkin gate [10]

In Fig.1, $D_a$ and $D_b$ are two single photon detectors. CNOT gate can make the polarization of the input photon rotate $90^0$, that is, $|H\rangle \to |V\rangle$ and $|V\rangle \to |H\rangle$. $HWP_1$ and $HWP_2$ are 22.5°-tilted half-wave plates. $HWP_3$ and $HWP_4$ are 67.5°-tilted half-wave plates. The functions of $HWP_1$ and $HWP_2$ are shown as below:

$$HWP_1, HWP_2 \to \begin{cases} |H\rangle \to \frac{1}{\sqrt{2}}(|H\rangle+|V\rangle) \\ |V\rangle \to \frac{1}{\sqrt{2}}(|H\rangle-|V\rangle) \end{cases}$$

The functions of $HWP_3$ and $HWP_4$ are shown as below:

$$HWP_3, HWP_4 \to \begin{cases} |H\rangle \to \frac{1}{\sqrt{2}}(|V\rangle-|H\rangle) \\ |V\rangle \to \frac{1}{\sqrt{2}}(|H\rangle+|V\rangle) \end{cases}$$

When the photons are transmitted through polarization beam splitters $PBS_1$ and $PBS_2$, the state is

$$|\psi_{1,2,3,4}\rangle = \beta_{1h}\beta_{2h}|H\rangle_2|H\rangle_3 + \beta_{1h}\beta_{2v}|H\rangle_2|V\rangle_4 + \beta_{1v}\beta_{2h}|V\rangle_1|H\rangle_3 + \beta_{1v}\beta_{2v}|V\rangle_1|V\rangle_4 \qquad (2)$$

If $|c_{in}\rangle = |1\rangle = |V\rangle$, the state of the photons transmitted through CNOT gates, $PBS_3$ and

$PBS_4$ is

$$|\psi_{5,6,7,8}\rangle = \beta_{1h}\beta_{2h}|V\rangle_5|V\rangle_6 + \beta_{1h}\beta_{2v}|V\rangle_5|H\rangle_7 + \beta_{1v}\beta_{2h}|H\rangle_8|V\rangle_6 + \beta_{1v}\beta_{2v}|H\rangle_8|H\rangle_7 \quad (3)$$

The state of photons transmitted through $HWP_1$, $HWP_2$, $HWP_3$ and $HWP_4$ is

$$\begin{aligned}|\psi_{5',6',7',8'}\rangle = \frac{1}{2}\big[&\beta_{1h}\beta_{2h}(|H\rangle_{5'}-|V\rangle_{5'})(|H\rangle_{6'}-|V\rangle_{6'}) + \beta_{1h}\beta_{2v}(|H\rangle_{5'}-|V\rangle_{5'})(|V\rangle_{7'}-|H\rangle_{7'}) \\ +&\beta_{1v}\beta_{2h}(|V\rangle_{8'}-|H\rangle_{8'})(|H\rangle_{6'}-|V\rangle_{6'}) + \beta_{1v}\beta_{2v}(|V\rangle_{8'}-|H\rangle_{8'})(|V\rangle_{7'}-|H\rangle_{7'})\big]\end{aligned} \quad (4)$$

The state of photons after $PBS_5$ and $PBS_6$ is

$$\begin{aligned}|\psi_{b_1,b_1',b_2,b_2'}\rangle = \frac{1}{2}\big[&\beta_{1h}\beta_{2h}(|H\rangle_{b_2}-|V\rangle_{b_2'})(|H\rangle_{b_1}-|V\rangle_{b_1'}) + \beta_{1h}\beta_{2v}(|H\rangle_{b_2}-|V\rangle_{b_2'})(|V\rangle_{b_1}-|H\rangle_{b_1'}) \\ +&\beta_{1v}\beta_{2h}(|V\rangle_{b_2}-|H\rangle_{b_2'})(|H\rangle_{b_1}-|V\rangle_{b_1'}) + \beta_{1v}\beta_{2v}(|V\rangle_{b_2}-|H\rangle_{b_2'})(|V\rangle_{b_1}-|H\rangle_{b_1'})\big]\end{aligned} \quad (5)$$

If no photons arrive at SPD $D_a$ and $D_b$, the output state is

$$|\psi_{out}\rangle = \beta_{1h}\beta_{2h}|H\rangle_{b_2}|H\rangle_{b_1} + \beta_{1h}\beta_{2v}|H\rangle_{b_2}|V\rangle_{b_1} + \beta_{1v}\beta_{2h}|V\rangle_{b_2}|H\rangle_{b_1} + \beta_{1v}\beta_{2v}|V\rangle_{b_2}|V\rangle_{b_1} \quad (6)$$

In this case, the quantum Fredkin gate is in "Cross" state. Otherwise, if $|c_{in}\rangle = |0\rangle = |H\rangle$ and no photon arrives at $D_a$ and $D_b$, the output state is

$$|\psi_{out}\rangle = \beta_{1h}\beta_{2h}|H\rangle_{b_1}|H\rangle_{b_2} + \beta_{1h}\beta_{2v}|H\rangle_{b_1}|V\rangle_{b_2} + \beta_{1v}\beta_{2h}|V\rangle_{b_1}|H\rangle_{b_2} + \beta_{1v}\beta_{2v}|V\rangle_{b_1}|V\rangle_{b_2} \quad (7)$$

In this case, the quantum Fredkin gate is in "Through" state. Let $\eta$ and $p_d$ denote the efficiency and the dark count of SPD, respectively. The probability $p_{11}$ with which no photon is counted by $D_a$ and $D_b$ is

$$\begin{aligned}p_{11} &= \frac{1}{4}\big[(1-\eta)^2 + 2(1-\eta)(1-p_d) + (1-p_d)^2\big] \\ &= \frac{1}{4}(2-\eta-p_d)^2\end{aligned} \quad (8)$$

Since $p_d \ll \eta$,

$$p_{11} \approx \frac{1}{4}(2-\eta)^2 \quad (9)$$

From (9) we know that the probability $p_{11}$ is decreased quickly with the detector efficiency $\eta$ increasing. That is, the lower the detector efficiency, the higher $p_{11}$. This is a strange result. Only when the quantum efficiency $\eta$ of SPD is 100% and the dark count isn't taken into account, is $p_{11} = \frac{1}{4}$ the success rate of the quantum Fredkin gate. Note that the success rate here means the probability with which the Fredkin gate is heralded as performing a successful operation. Actually, the probability includes false heralding probability because of low-efficient SPD. So, the fidelity of the gate is smaller than 1.

**2.2 The success rate of heralded linear optical quantum Fredkin gate under non-ideal SPS**

Nowadays, there is no ideal single photon source which can be applied to practical quantum communication system. Attenuated laser is often used to taken as quasi-single photon source in which photon number of a pulse is poisson distributed. By using this kind of source and the scheme shown in Fig.1 the success rate could be further worsen. We analyze this scenario briefly.

Firstly, we assume that there are two photons in each input pulse. The input states at $a_{1in}$ and $a_{2in}$ are $|\psi_{a_1}\rangle = (\beta_{1h}|H\rangle_{a_1} + \beta_{1v}|V\rangle_{a_1})(\beta_{1h}|H\rangle_{a_1} + \beta_{1v}|V\rangle_{a_1})$ and $|\psi_{a_2}\rangle = (\beta_{2h}|H\rangle_{a_2} + \beta_{2v}|V\rangle_{a_2})(\beta_{2h}|H\rangle_{a_2} + \beta_{2v}|V\rangle_{a_2})$. So, the joint input state is

$$\begin{aligned}|\psi_{in}\rangle = &\beta_{1h}^2\beta_{2h}^2|HH\rangle_{a_1}|HH\rangle_{a_2} + \beta_{1h}^2\beta_{2h}\beta_{2v}|HH\rangle_{a_1}|HV\rangle_{a_2} + \beta_{1h}^2\beta_{2h}\beta_{2v}|HH\rangle_{a_1}|VH\rangle_{a_2}\\ &+ \beta_{1h}^2\beta_{2v}^2|HH\rangle_{a_1}|VV\rangle_{a_2} + \beta_{1h}\beta_{1v}\beta_{2h}^2|HV\rangle_{a_1}|HH\rangle_{a_2} + \beta_{1h}\beta_{1v}\beta_{2h}\beta_{2v}|HV\rangle_{a_1}|HV\rangle_{a_2}\\ &+ \beta_{1h}\beta_{1v}\beta_{2h}\beta_{2v}|HV\rangle_{a_1}|VH\rangle_{a_2} + \beta_{1h}\beta_{1v}\beta_{2v}^2|HV\rangle_{a_1}|VV\rangle_{a_2} + \beta_{1h}\beta_{1v}\beta_{2h}^2|VH\rangle_{a_1}|HH\rangle_{a_2}\\ &+ \beta_{1h}\beta_{1v}\beta_{2h}\beta_{2v}|VH\rangle_{a_1}|HV\rangle_{a_2} + \beta_{1h}\beta_{1v}\beta_{2h}\beta_{2v}|VH\rangle_{a_1}|VH\rangle_{a_2} + \beta_{1h}\beta_{1v}\beta_{2v}^2|VH\rangle_{a_1}|VV\rangle_{a_2}\\ &+ \beta_{1v}^2\beta_{2h}^2|VV\rangle_{a_1}|HH\rangle_{a_2} + \beta_{1v}^2\beta_{2h}\beta_{2v}|VV\rangle_{a_1}|HV\rangle_{a_2} + \beta_{1v}^2\beta_{2h}\beta_{2v}|VV\rangle_{a_1}|VH\rangle_{a_2}\\ &+ \beta_{1v}^2\beta_{2v}^2|VV\rangle_{a_1}|VV\rangle_{a_2}\end{aligned}$$

(10)

If $|c_{in}\rangle = |1\rangle = |V\rangle$, then the output state is

$$\begin{aligned}|\psi_{b_1,b_1',b_2,b_2'}\rangle = \frac{1}{4}\Big[&\beta_{1h}^2\beta_{2h}^2(|H\rangle_{b_2}-|V\rangle_{b_2'})(|H\rangle_{b_2}-|V\rangle_{b_2'})(|H\rangle_{b_1}-|V\rangle_{b_1'})(|H\rangle_{b_1}-|V\rangle_{b_1'})\\ &+ \beta_{1h}^2\beta_{2h}\beta_{2v}(|H\rangle_{b_2}-|V\rangle_{b_2'})(|H\rangle_{b_2}-|V\rangle_{b_2'})(|H\rangle_{b_1}-|V\rangle_{b_1'})(|V\rangle_{b_1}-|H\rangle_{b_1'})\end{aligned}$$

$$+ \beta_{1h}^2\beta_{2h}\beta_{2v}\left(|H\rangle_{b_2}-|V\rangle_{b_2'}\right)\left(|H\rangle_{b_2}-|V\rangle_{b_2'}\right)\left(|V\rangle_{b_1}-|H\rangle_{b_1'}\right)\left(|H\rangle_{b_1}-|V\rangle_{b_1'}\right)$$

$$+ \beta_{1h}^2\beta_{2v}^2\left(|H\rangle_{b_2}-|V\rangle_{b_2'}\right)\left(|H\rangle_{b_2}-|V\rangle_{b_2'}\right)\left(|V\rangle_{b_1}-|H\rangle_{b_1'}\right)\left(|V\rangle_{b_1}-|H\rangle_{b_1'}\right)$$

$$+ \beta_{1h}\beta_{1v}\beta_{2h}^2\left(|H\rangle_{b_2}-|V\rangle_{b_2'}\right)\left(|V\rangle_{b_2}-|H\rangle_{b_2'}\right)\left(|H\rangle_{b_1}-|V\rangle_{b_1'}\right)\left(|H\rangle_{b_1}-|V\rangle_{b_1'}\right)$$

$$+ \beta_{1h}\beta_{1v}\beta_{2h}\beta_{2v}\left(|H\rangle_{b_2}-|V\rangle_{b_2'}\right)\left(|V\rangle_{b_2}-|H\rangle_{b_2'}\right)\left(|H\rangle_{b_1}-|V\rangle_{b_1'}\right)\left(|V\rangle_{b_1}-|H\rangle_{b_1'}\right)$$

$$+ \beta_{1h}\beta_{1v}\beta_{2h}\beta_{2v}\left(|H\rangle_{b_2}-|V\rangle_{b_2'}\right)\left(|V\rangle_{b_2}-|H\rangle_{b_2'}\right)\left(|V\rangle_{b_1}-|H\rangle_{b_1'}\right)\left(|H\rangle_{b_1}-|V\rangle_{b_1'}\right)$$

$$+ \beta_{1h}\beta_{1v}\beta_{2v}^2\left(|H\rangle_{b_2}-|V\rangle_{b_2'}\right)\left(|V\rangle_{b_2}-|H\rangle_{b_2'}\right)\left(|V\rangle_{b_1}-|H\rangle_{b_1'}\right)\left(|V\rangle_{b_1}-|H\rangle_{b_1'}\right)$$

$$+ \beta_{1h}\beta_{1v}\beta_{2h}^2\left(|V\rangle_{b_2}-|H\rangle_{b_2'}\right)\left(|H\rangle_{b_2}-|V\rangle_{b_2'}\right)\left(|H\rangle_{b_1}-|V\rangle_{b_1'}\right)\left(|H\rangle_{b_1}-|V\rangle_{b_1'}\right)$$

$$+ \beta_{1h}\beta_{1v}\beta_{2h}\beta_{2v}\left(|V\rangle_{b_2}-|H\rangle_{b_2'}\right)\left(|H\rangle_{b_2}-|V\rangle_{b_2'}\right)\left(|H\rangle_{b_1}-|V\rangle_{b_1'}\right)\left(|V\rangle_{b_1}-|H\rangle_{b_1'}\right)$$

$$+ \beta_{1h}\beta_{1v}\beta_{2h}\beta_{2v}\left(|V\rangle_{b_2}-|H\rangle_{b_2'}\right)\left(|H\rangle_{b_2}-|V\rangle_{b_2'}\right)\left(|V\rangle_{b_1}-|H\rangle_{b_1'}\right)\left(|H\rangle_{b_1}-|V\rangle_{b_1'}\right)$$

$$+ \beta_{1h}\beta_{1v}\beta_{2v}^2\left(|V\rangle_{b_2}-|H\rangle_{b_2'}\right)\left(|H\rangle_{b_2}-|V\rangle_{b_2'}\right)\left(|V\rangle_{b_1}-|H\rangle_{b_1'}\right)\left(|V\rangle_{b_1}-|H\rangle_{b_1'}\right)$$

$$+ \beta_{1v}^2\beta_{2h}^2\left(|V\rangle_{b_2}-|H\rangle_{b_2'}\right)\left(|V\rangle_{b_2}-|H\rangle_{b_2'}\right)\left(|H\rangle_{b_1}-|V\rangle_{b_1'}\right)\left(|H\rangle_{b_1}-|V\rangle_{b_1'}\right)$$

$$+ \beta_{1v}^2\beta_{2h}\beta_{2v}\left(|V\rangle_{b_2}-|H\rangle_{b_2'}\right)\left(|V\rangle_{b_2}-|H\rangle_{b_2'}\right)\left(|H\rangle_{b_1}-|V\rangle_{b_1'}\right)\left(|V\rangle_{b_1}-|H\rangle_{b_1'}\right)$$

$$+ \beta_{1v}^2\beta_{2h}\beta_{2v}\left(|V\rangle_{b_2}-|H\rangle_{b_2'}\right)\left(|V\rangle_{b_2}-|H\rangle_{b_2'}\right)\left(|V\rangle_{b_1}-|H\rangle_{b_1'}\right)\left(|H\rangle_{b_1}-|V\rangle_{b_1'}\right)$$

$$+ \beta_{1v}^2\beta_{2v}^2\left(|V\rangle_{b_2}-|H\rangle_{b_2'}\right)\left(|V\rangle_{b_2}-|H\rangle_{b_2'}\right)\left(|V\rangle_{b_1}-|H\rangle_{b_1'}\right)\left(|V\rangle_{b_1}-|H\rangle_{b_1'}\right)$$

(11)

If no photon arrives at $D_a$ or $D_b$, the output state can be given as

$$|\psi_{out}\rangle = \frac{1}{4}\Big[\beta_{1h}^2\beta_{2h}^2|HH\rangle_{b_2}|HH\rangle_{b_1} + \beta_{1h}^2\beta_{2h}\beta_{2v}|HH\rangle_{b_2}|HV\rangle_{b_1} + \beta_{1h}^2\beta_{2h}\beta_{2v}|HH\rangle_{b_2}|VH\rangle_{b_1}$$
$$+ \beta_{1h}^2\beta_{2v}^2|HH\rangle_{b_2}|VV\rangle_{b_1} + \beta_{1h}\beta_{1v}\beta_{2h}^2|HV\rangle_{b_2}|HH\rangle_{b_1} + \beta_{1h}\beta_{1v}\beta_{2h}\beta_{2v}|HV\rangle_{b_2}|HV\rangle_{b_1}$$
$$+ \beta_{1h}\beta_{1v}\beta_{2h}\beta_{2v}|HV\rangle_{b_2}|VH\rangle_{b_1} + \beta_{1h}\beta_{1v}\beta_{2v}^2|HV\rangle_{b_2}|VV\rangle_{b_1} + \beta_{1h}\beta_{1v}\beta_{2h}^2|VH\rangle_{b_2}|HH\rangle_{b_1}$$
$$+ \beta_{1h}\beta_{1v}\beta_{2h}\beta_{2v}|VH\rangle_{b_2}|HV\rangle_{b_1} + \beta_{1h}\beta_{1v}\beta_{2h}\beta_{2v}|VH\rangle_{b_2}|VH\rangle_{b_1} + \beta_{1h}\beta_{1v}\beta_{2v}^2|VH\rangle_{b_2}|VV\rangle_{b_1}$$
$$+ \beta_{1v}^2\beta_{2h}^2|VV\rangle_{b_2}|HH\rangle_{b_1} + \beta_{1v}^2\beta_{2h}\beta_{2v}|VV\rangle_{b_2}|HV\rangle_{b_1} + \beta_{1v}^2\beta_{2h}\beta_{2v}|VV\rangle_{b_2}|VH\rangle_{b_1}$$
$$+ \beta_{1v}^2\beta_{2v}^2|VV\rangle_{b_2}|VV\rangle_{b_1}\Big]$$

(12)

In this case, the quantum Fedkin gate is in "Cross" state. If $|c_{in}\rangle = |0\rangle = |H\rangle$ and no photon arrives at $D_a$ or $D_b$, then the output state is as the same as the input state shown as (10). In this case, the quantum Fredkin gate is in "through" state.

We also obtain that the probability, $p_{22}$, with which no photon arrives at $D_a$ or $D_b$ (the quantum Fredkin gate is in "cross" or "through" state), can be given as

$$p_{22} = \frac{1}{16}\left[(1-p_d)^2 + 4(1-p_d)(1-\eta) + 2(1-p_d)(1-\eta)^2 \right. \\ \left. + 4(1-\eta)^2 + 4(1-\eta)^3 + (1-\eta)^4\right] \tag{13}$$

Secondly, if the number of photons in a pulse at $a_{1in}$ is $m$ and the number of photons in a pulse at $a_{2in}$ is $n$, then the probability, $p_{mn}$, with which no photon arrives at $D_a$ or $D_b$, is

$$p_{mn} = \frac{1}{2^{m+n}}\left[(1-p_d)^2 + (1-p_d)\sum_{m_1=1}^{m} C_m^{m_1}(1-\eta_d)^{m_1} + (1-p_d)\sum_{n_1=1}^{n} C_n^{n_1}(1-\eta_d)^{n_1} \right. \\ \left. + \sum_{m_1=1}^{m}\sum_{n_1=1}^{n} C_m^{m_1} C_n^{n_1}(1-\eta_d)^{m_1}(1-\eta_d)^{n_1}\right] \tag{14}$$

We assume the photon number in the input pulses at $a_{1in}$ is poisson distributed with mean $\mu_1$, the photon number in the input pulses at $a_{2in}$ is poisson distributed with mean $\mu_2$, and the photon numbers of them are independently distributed. So under non-ideal SPS and low efficient SPD the probability $p$, with which no photon arrives at $D_a$ or $D_b$, is

$$p = \sum_{m=0}^{+\infty}\sum_{n=0}^{+\infty} p_{mn} \cdot \frac{\mu_1^m}{m!} \cdot \frac{\mu_2^n}{n!} e^{-(\mu_1+\mu_2)} - (1-p_d)^2 e^{-(\mu_1+\mu_2)} \\ = \sum_{m=0}^{+\infty}\sum_{n=0}^{+\infty} \frac{1}{2^{m+n}}\left[(1-p_d)^2 + (1-p_d)\sum_{m_1=1}^{m} C_m^{m_1}(1-\eta)^{m_1} + (1-p_d)\sum_{n_1=1}^{n} C_n^{n_1}(1-\eta)^{n_1} \right. \\ \left. + \sum_{m_1=1}^{m}\sum_{n_1=1}^{n} C_m^{m_1} C_m^{m_1}(1-\eta)^{m_1}(1-\eta)^{m_1}\right] \cdot \frac{\mu_1^m}{m!} \cdot \frac{\mu_2^n}{n!} e^{-(\mu_1+\mu_2)} - (1-p_d)^2 e^{-(\mu_1+\mu_2)} \tag{15}$$

In equation (15), we define $C_0^k = 0, k > 0$. From (13)-(15), we can obtain that the lower the detector efficiency $\eta$ is, the higher the probability $p$ is. This is also a strange result.

## 3 Modified heralded linear optical quantum Fredkin gate

### 3.1 Linear optical quantum Fredkin gate with modified heralding circuit

We modify the heralding circuit of the quantum Fredkin gate, as shown in Fig.2. In this heralding scheme, the photons at $b_1'$ are transmitted through two paths: the photons with horizontal polarization pass through $PBS_7$ and $PBS_9$; the photons with vertical polarization pass through $PBS_7$, delay line $DL_1$ and $PBS_9$. The photons at $b_2'$ are also transmitted through two paths: the photons with horizontal polarization pass through $PBS_8$ and $PBS_{10}$; the photons with vertical polarization pass through $PBS_8$, delay line $DL_2$ and $PBS_{10}$.

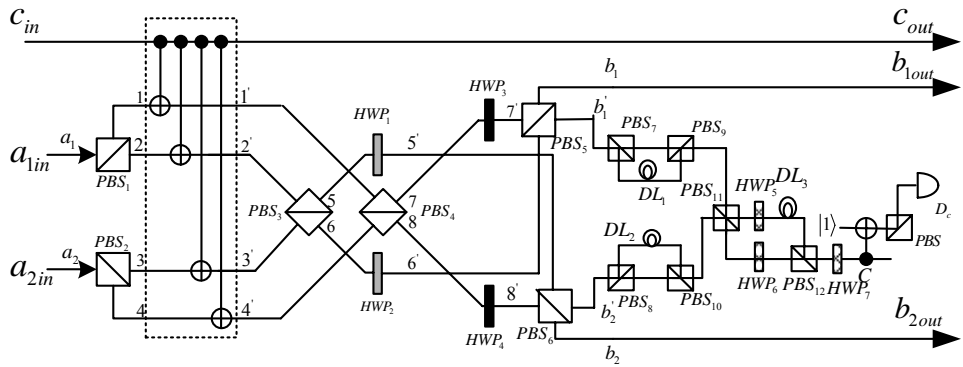

Fig.2. Modified heralded quantum Fredkin gate

In Fig.2, the time delay of $DL_1$ is equal to the one of $DL_2$ and longer than two times of the time delay of $DL_3$. The half wave plates $HWP_5$, $HWP_6$ and $HWP_7$ are designed to rotate the polarization direction by $90°$ and they can work or not according to control signal (this can be implemented by dynamic polarization controller). The application of time delay lines and HWPs can make the photons be detected by only one SPD in sequence. There are four scenarios when photons pass through $PBS_{11}$: (1) the horizontally polarized photons from $b_1'$ pass through $PBS_{11}$, $HWP_6$ which doesn't work, $PBS_{12}$ and $HWP_7$ which works by rotating the horizontal polarization to vertical polarization; (2) the horizontally polarized photons from $b_2'$ pass through $PBS_{11}$, $HWP_5$ which works by rotating the horizontal polarization to vertical polarization, $DL_3$, $PBS_{12}$ and $HWP_7$ which doesn't work; (3) the vertically polarized photons from $b_2'$ pass through $PBS_{11}$, $HWP_6$ which works by rotating the vertical polarization to horizontal polarization, $PBS_{12}$ and $HWP_7$ which works by rotating the horizontal polarization to vertical polarization; (4) the vertically polarized photons from $b_1'$ pass through $PBS_{11}$, $HWP_5$ which doesn't work, $DL_3$, $PBS_{12}$ and $HWP_7$ which doesn't work. So, all the photons which can arrive at the output port of $HWP_7$ must be vertically polarized. That is, they are qubits denoted by $|V\rangle$ or $|1\rangle$. Then, these photons are connected to the control port of a CNOT gate with $|1\rangle$ as target qubit. So, if one or more photons arrive at $b_1'$ or $b_2'$, the detector $D_c$ will be not ticked. Otherwise, if no photon arrives at $b_1'$ or $b_2'$, the detector $D_c$ will be ticked. So the detector $D_c$ can be used to imply the success of the quantum Fredkin gate. The Fredkin gate succeeds if $D_c$ is ticked. The higher the efficiency of $D_c$, the higher the efficiency of heralding.

If the photon number in a pulse at $a_{1in}$ is $m$ and the photon number in a pulse at $a_{2in}$ is $n$, then the probability $p_{mn}$, with which the detector $D_c$ is ticked, can be given as

$$p_{mn} = \frac{1}{2^{m+n}}\eta + \left(1 - \frac{1}{2^{m+n}}\right)p_d, \qquad m,n = 0,1,2,\ldots, but\ m = n \neq 0 \tag{16}$$

So the mean count rate $p_{new}$ can be given as

$$p_{new} = \sum_{m=0}^{+\infty}\sum_{n=0}^{+\infty} p_{mn} \cdot \frac{\mu_1^m}{m!} \cdot \frac{\mu_2^n}{n!} e^{-(\mu_1+\mu_2)} - \eta e^{-(\mu_1+\mu_2)}$$

$$= \sum_{m=0}^{+\infty}\sum_{n=0}^{+\infty}\left[\frac{1}{2^{m+n}}\eta + \left(1 - \frac{1}{2^{m+n}}\right)p_d\right] \cdot \frac{\mu_1^m}{m!} \cdot \frac{\mu_2^n}{n!} e^{-(\mu_1+\mu_2)} - \eta e^{-(\mu_1+\mu_2)}$$

$$= \sum_{m=0}^{+\infty} \sum_{n=0}^{+\infty} \left[ (\eta - p_d) \cdot \frac{\left(\mu_1/2\right)^m}{m!} \cdot e^{-\mu_1} \cdot \frac{\left(\mu_2/2\right)^n}{n!} \cdot e^{-\mu_2} + p_d \cdot \frac{\mu_1^m}{m!} \cdot \frac{\mu_2^n}{n!} e^{-(\mu_1+\mu_2)} \right]$$

$$-\eta e^{-(\mu_1+\mu_2)}$$

$$= (\eta - p_d) e^{-\frac{\mu_1+\mu_2}{2}} + p_d - \eta e^{-(\mu_1+\mu_2)} \tag{17}$$

Let $\mu_1 = \mu_2 = \mu$. Since $p_d \ll \eta$, the mean count rate $p_{new}$ can be approximately given as

$$p_{new} \approx \eta \left( e^{-\mu} - e^{-2\mu} \right) \tag{18}$$

In this case, $p_{new}$ is just as the success rate of quantum Fredkin gate. The probability $p_{new}$ increases linearly with the detector efficiency $\eta$, as shown in Figure 3. The higher the average photon number is, the higher the success rate is. When $\mu = 0.1$ and detector is ideal (detection efficiency is 100%), the success rate of the Fredkin gate is about 0.085.

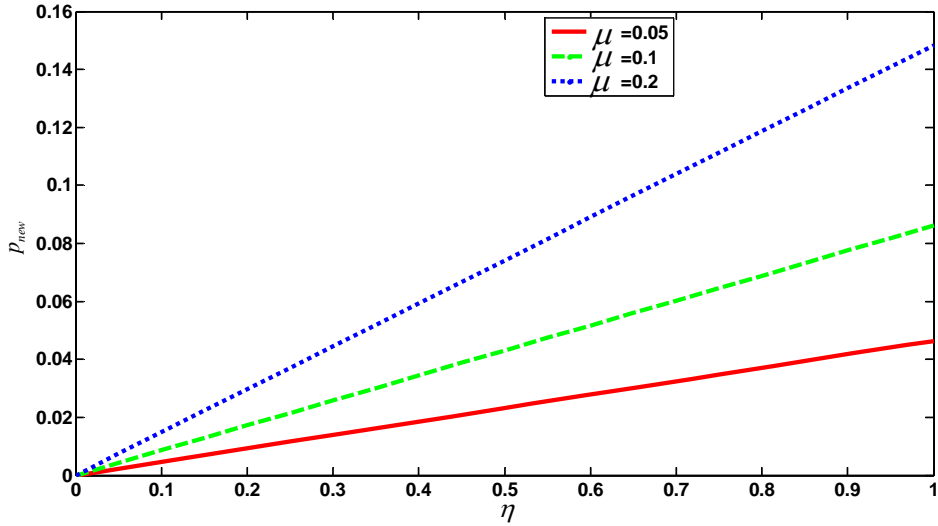

**Fig. 3.** The success rate, $p_{new}$, of modified quantum Fredkin gate

In Figure 2, the CNOT gate can be implemented without entangled ancilla[19], or can be implemented by integrated photonics[20].

### 3.2 Quantum Benes network built with quantum Fredkin gate

Based on the modified heralded quantum Fredkin gate shown in Fig.2, quantum switch network can be built. The Fredkin gate is the basic switch unit with three inputs, three outputs and one flag, as shown in Fig.4a. A quantum Benes network can be shown in Fig.4b. The symbol "&" denotes a six-qubit quantum "AND" gate, which can be created using conditional NOT with six control qubits or combination of several CNOT gates.

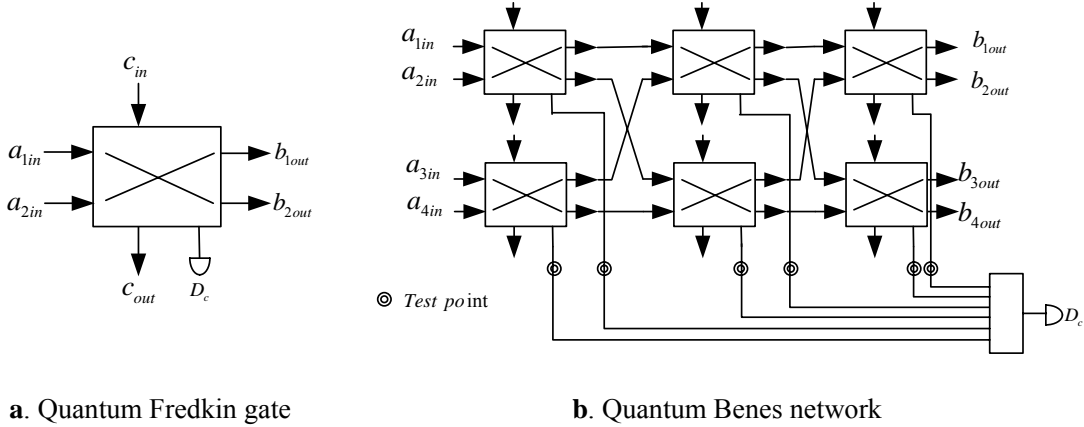

    **a**. Quantum Fredkin gate             **b**. Quantum Benes network

**Fig.4.** Quantum Benes network built with quantum Fredkin gate

In Fig.4, the quantum Benes network needs only one SPD. We also present the test points. So whether each Fredkin gate works well or not can be checked easily.

## 4 Conclusion

In this paper, we propose a linear optical quantum Fredkin gate in which a new heralding quantum circuit with one SPD is applied and an auxiliary qubit $|1\rangle$ is added. In this scheme, the success rate of the quantum Fredkin gate is proportional to the efficiency of the heralding detector. With this improved heralding scheme the gate can be easily used to build large-scale quantum switching network. It doesn't require interferometer and entangled auxiliary photons. It should be mentioned that according to the gate fidelity defined as [21], the fidelity of the gate shown in Fig. 1 is $1/(2-\eta)^2$ for $p_d = 0$ and ideal single photon source, while the fidelity of the gate shown in Fig. 2 is nearly 1. Our further works are to make experiment and to research how to implement it by integrated optics technology.

# Acknowledgements

This work was supported by the National Natural Science Foundation of China No.61372076 and No. 61301171, the 111 Project no.B08038, and the Fundamental Research Funds for the Central Universities no. K5051301059 and K5051201021.